\documentclass[sigconf]{acmart}

\usepackage{balance}
\usepackage{amsmath,amsfonts}
\usepackage{algorithmic}
\usepackage{color}
\usepackage{enumitem}
\usepackage{fancyhdr}
\usepackage{graphicx}
\usepackage[utf8]{inputenc}
\usepackage{listings}
\usepackage{makecell}
\usepackage{mathtools}
\usepackage{microtype}
\usepackage{multirow}
\usepackage{pifont}
\usepackage{tablefootnote}
\usepackage{textcomp}
\usepackage[flushleft]{threeparttable}
\usepackage[normalem]{ulem}
\usepackage{xcolor}
\useunder{\uline}{\ul}{}

\definecolor{dark}{rgb}{0.2,0.2,0}
\definecolor{lightred}{rgb}{1,0.3,0.3}
\definecolor{darkyellow}{rgb}{0.4,0.4,0}
\definecolor{mediumgreen}{rgb}{0,0.6,0}
\definecolor{darkgreen}{rgb}{0.1,0.5,0}
\definecolor{lightblue}{rgb}{0.6,0.7,1}
\definecolor{darkblue}{rgb}{0.0,0.0,0.5}
\definecolor{darkred}{rgb}{0.25,0.0,0.0}
\definecolor{listb}{rgb}{1.0,1.0,0.95}

\lstdefinestyle{mystyle}{
  breakatwhitespace=false,         
  breaklines=true,                 
  captionpos=b,     
  emph={include,define,int,unsigned,for,do,while,return,auto,const,using,template,namespace,constexpr,long,class,Enc,void,vector,true,bool,else},
  frame = single,
  emphstyle=\bf,
  keepspaces=true,       
  mathescape=true,
  numbers=left,                    
  numberstyle=\footnotesize, 
  showspaces=false,                
  showstringspaces=false,
  showtabs=false,                  
  tabsize=2,
  basicstyle=\ttfamily\footnotesize,
  keywordstyle=\color{violet}\bf,
  keywords={SecureInt, SecureUint, SecureMod, SecInt, SecUint, SecureBool},
  xleftmargin=2em,
  numbersep=1em,
  morecomment=[s][\color{blue}]{"}{"},
  moredelim=**[is][\color{red}]{@}{@} 
}

\lstdefinestyle{mystyleb}{
  backgroundcolor=\color{listb},   commentstyle=\color{codegreen},
  breakatwhitespace=false,         
  breaklines=true,                 
  captionpos=b,     
  emph={include,define,int,unsigned,for,do,while,return,auto,const,using,template,namespace,constexpr,long,class,Enc,void,vector,true,bool,else},
  frame = single,
  emphstyle=\bf,
  keepspaces=true,       
  mathescape=true,
  numbers=left,                    
  numberstyle=\footnotesize, 
  showspaces=false,                
  showstringspaces=false,
  showtabs=false,                  
  tabsize=2,
  basicstyle=\ttfamily\footnotesize,
  keywordstyle=\color{darkblue}\bf,
  keywords={SecureInt, SecureUint, SecureMod, SecInt, SecureBool},
  xleftmargin=2em,
  numbersep=1em,
  morecomment=[s][\color{blue}]{"}{"},
  moredelim=**[is][\color{red}]{@}{@} 
}

\newcommand{\draftcom}[1]{}
\newcommand{\eee}{{{E3}}}
\newcommand{\ignore}[1]{}

\newcommand{\secint}{{\tt{}SecureInt}}
\newcommand{\secuint}{{\tt{}SecureUint}}
\newcommand{\secmod}{{\tt{}SecureMod}}

\newcommand{\secbool}{{\tt SecureBool}}

\newcommand{\subsubsubsection}[1]{\noindent\textbf{#1}:}
\newcommand{\innersection}[1]{\subsubsubsection{#1}}

\newcommand{\replace}[2][]{{#2}} 

\DeclarePairedDelimiter{\ceil}{\lceil}{\rceil}
\DeclarePairedDelimiter{\floor}{\lfloor}{\rfloor}

\AtBeginDocument{%
  \providecommand\BibTeX{{%
    \normalfont B\kern-0.5em{\scshape i\kern-0.25em b}\kern-0.8em\TeX}}}

\copyrightyear{2022}
\acmYear{2022}
\setcopyright{acmlicensed}\acmConference[ICCAD '22]{IEEE/ACM International Conference on Computer-Aided Design}{October 30-November 3, 2022}{San Diego, CA, USA}
\acmBooktitle{IEEE/ACM International Conference on Computer-Aided Design (ICCAD '22), October 30-November 3, 2022, San Diego, CA, USA}
\acmPrice{15.00}
\acmDOI{10.1145/3508352.3549415}
\acmISBN{978-1-4503-9217-4/22/10}

\begin{document}

\title[Accelerating Fully Homomorphic Encryption by Bridging Modular and Bit-Level Arithmetic]
{Accelerating Fully Homomorphic Encryption\\by Bridging Modular and Bit-Level Arithmetic}


\author{Eduardo Chielle, Oleg Mazonka, Homer Gamil, and Michail Maniatakos}
\affiliation{
    \institution{Center for Cyber Security, New York University Abu Dhabi, Abu Dhabi, UAE}
    \country{}
}
\email{{eduardo.chielle, om22, homer.g, michail.maniatakos}@nyu.edu}







\begin{abstract}
The dramatic increase of data breaches in modern computing platforms has emphasized that access control is not sufficient to protect sensitive user data. Recent advances in cryptography allow end-to-end processing of encrypted data without the need for decryption using Fully Homomorphic Encryption (FHE).
Such computation however, is still orders of magnitude slower than direct (unencrypted) computation. Depending on the underlying cryptographic scheme, FHE schemes can work natively either at bit-level using Boolean circuits, or over integers using modular arithmetic. Operations on integers are limited to addition/subtraction and multiplication. On the other hand, bit-level arithmetic is much more comprehensive allowing more operations, such as comparison and division. While modular arithmetic can emulate bit-level computation, there is a significant cost in performance.
In this work, we propose a novel method, dubbed \emph{bridging}, that blends faster and restricted modular computation with slower and comprehensive bit-level computation, making them both usable within the same application and with the same cryptographic scheme instantiation. We introduce and open source C++ types representing the two distinct arithmetic modes, offering the possibility to convert from one to the other.
Experimental results show that bridging modular and bit-level arithmetic computation can lead to 1-2 orders of magnitude performance improvement for tested synthetic benchmarks, as well as  one real-world FHE application: a genotype imputation case study.
\end{abstract}



\maketitle

\section{Introduction}

With the ever increasing rates of data generation, digital information is becoming extremely valuable over time. As a result, certain types of attacks have emerged, focusing to capitalize on this paradigm shift. To overcome the aforementioned threats, the use of cryptography has been until now the defacto technological security measure to protect against data leakage. Even though accepted standards such as AES have been successful in protecting data-in-transit and data-at-rest, they fail to provide protection towards data-in-use. Hardware solutions \cite{costan2016sgx-explained,trustzone} attempt to provide confidentiality and integrity on sensitive computations, however, a number of attacks \cite{yanga2, hely2012malicious, jin2012exposing, xiao2016hardware, karri2010trustworthy, tsoutsos2013fabrication,adee2008hunt, gross2018ending} including Spectre \cite{Kocher2018spectre}, Meltdown \cite{Lipp2018meltdown}, and Load Value Injection \cite{vanbulck2020lvi}, have raised questions about their effectiveness. Data are eventually decrypted before entering the processor pipeline, and therefore leakage is still possible as recently reported \cite{chen2018sgxpectre}.

A solution to the problem of protecting data-in-use is Fully Homomorphic Encryption (FHE), as it allows unconstrained computations on ciphertexts. Numerous FHE schemes have been developed, namely BGV~\cite{BGV_ref}, BFV~\cite{fan2012somewhatmisc}, CKKS~\cite{CKKS_ref}, GSW~\cite{GSW_ref}. 
GSW exposes homomorphic Boolean gates and programming computation can be constructed bottom-up.
Such bit-level arithmetic is universal, but requires many Boolean logic operations.
Other schemes, like BGV/BFV, operate directly on integers using modular arithmetic supporting homomorphic addition, subtraction, and multiplication.
At the level of integers however, the operations that can be applied on such ciphertexts are limited to those supported by the encryption scheme.
If an application requires another type of operation like an integer division or comparison, then all computation must be evaluated in bit-level arithmetic, by using arithmetic modulo 2 or algebraic expressions for Boolean gates.



In 2009, FHE received criticism on its high computational overheads, which made it challenging at the time to employ in practical applications. Improvements have since been performed on various fronts: 1) Improvement of implementation of FHE libraries. For example, HElib was recently overhauled to achieve almost $75 \times$ performance improvement \cite{cryptoeprint:2018:244}; 2) Hardware acceleration. The developers of nuFHE \cite{nuFHE} report about $100 \times$ acceleration over the software-based implementation of TFHE using GPU hardware. Meanwhile, F1, an ASIC accelerator for the BGV encryption scheme, outperforms software implementations by $5,400 \times$ \cite{f1}; 3) Dedicated programming frameworks. These frameworks allow users to express programming intent directly in a general-purpose programming language, such as Go, Python, C++.
For example, Lattigo~\cite{lattigop} implements  BFV and CKKS schemes in the Go programming language.
PyFHE, PySEAL and Pyfhel~\cite{pyfhel} are frameworks providing Python interface to FHE operations.
\eee\ framework \cite{e3eprint} offers custom C++ data types that can abstract the complexity of FHE library operations. As we observe, native modular arithmetic must be used for logic bit operations; hence program variables must be represented as a sequences of encrypted bits. In such case, \textit{all} programming operations are performed by evaluating Boolean circuits working on encrypted bits - ciphertexts. This transformation makes all computational operations slow, including addition and multiplication.

\innersection{Our contribution}
In this work, our proposal is to leverage the properties of both (1) bit-level arithmetic (universal, but slower) and (2) modular arithmetic (faster, but not universal) within the same algorithm expressed as a C++ program with custom data types, which allows much better performance of general-purpose programs processing encrypted data. 
Towards that end, we extend the E3 programming framework \cite{e3eprint} and implement \emph{Bridging}, which enables the mixing of different arithmetic abstractions in the same C++ program that processes FHE data.
The benefit of using our method is a significant application performance improvement, as we demonstrate with our experiments.

\section{Preliminaries}\label{s:preliminaries}

\subsection{Homomorphic Encryption}

Homomorphic encryption is a special type of encryption that allows meaningful operations in the encrypted domain. Within homomorphic encryption, there are several types of encryption schemes. Partially Homomorphic Encryption (PHE) schemes support only one homomorphic operation over ciphertexts, usually either addition or multiplication. FHE schemes support two orthogonal operations, usually addition and multiplication, theoretically allowing Turing complete computation.

Operations on ciphertexts without decryption are possible using homomorphic functions. Suppose the homomorphic function F over ciphertexts $c_x$ and $c_y$ corresponds to plaintext operation $f$.
Let $f(m_x,m_y)$ denote a two-argument function on plaintexts $m_x$ and $m_y$; $\text{E}_k(m,r)$ denotes a probabilistic encryption function generated by a random sequence $k$ (key) that maps a plaintext $m$ to a set of ciphertexts $c$ depending on a probabilistic parameter $r$, and $\text{D}_k(c)$ denotes a deterministic decryption function corresponding to $\text{E}_k$ that maps ciphertext $c$ to the corresponding plaintext $m$. Then,
F is defined by the homomorphism of surjective
$\text{D}_k$:
\begin{equation*}
\text{D}_k\big(\text{F}(c_x,c_y)\big) = 
f\big(\text{D}_k(c_x), \text{D}_k(c_y)\big),
\end{equation*}
which converts into the
explicit composition over ciphertexts:
\begin{equation*}\label{e:chop}
\text{F}(c_x,c_y) = 
\text{E}_k\Big(f\big(\text{D}_k(c_x), \text{D}_k(c_y)\big),\text{H}(c_x,c_y)\Big)
\end{equation*}
where H is an arbitrary function generating the randomness value $r$ from the input ciphertexts. This expression establishes a functional requirement for F, which does not decrypt nor re-encrypt the data.




\subsection{Data-oblivious programming}

{\it Data-oblivious computation} is a type of computation where the input data does not influence the behavior of the program. Under ``behavior'' we imply execution branching or memory access. 
Operations performed on encrypted data have to be in a data-oblivious form since ciphertexts are never decrypted, therefore the program is not capable of making decisions based on their plaintext values. In some fields, this property is also termed as {\it constant-time programming}, and has also been used for protecting against side-channel analysis.


Consider a simple function in Listing~\ref{list:fibo}. The iterations are interrupted once the index {\tt i} reaches the input value \texttt{in}. Assume that computation is now protected so the inputs and outputs must remain encrypted. Since variable {\tt in} cannot participate in any decisions to interrupt iterations, it must be processed in a data-oblivious way. We replace the interrupt condition with a multiplexer accumulator {\tt r} and introduce a fixed number of iterations {\tt max\_iter} not depending on the input, as shown in Listing~\ref{list:fibdo}. The number of iterations should exceed the input index, otherwise the accumulator will not reach the input index and will not get updated. The program shown in Listing~\ref{list:fibdo} works in constant time regardless of the input.

\begin{figure}
\begin{minipage}{\linewidth}
\begin{lstlisting}[language=C++, caption={Simple Fibonacci function.
% \hspace{1cm} \mbox{}
}, style=mystyle, label=list:fibo, 
% xrightmargin=-0.12\linewidth,
% linewidth=0.96\linewidth
]
int fibonacci(int in)
{
    int i=0, a=0, b=1;
    while( i++ != in )
    {
        std::swap(a,b);
        a += b;
    }
    return a;
}
\end{lstlisting}
\end{minipage}
\vspace{-0.1in}
\end{figure}

\begin{figure}
\begin{minipage}{\linewidth}
\begin{lstlisting}[language=C++, caption={Data-oblivious Fibonacci function.
% \hspace{-0.05cm} \mbox{}
}, style=mystyle, label=list:fibdo,
% xrightmargin=-0.12\linewidth,
%linewidth=0.96\linewidth
]
int fibonacci(int in)
{
    int i=0, a=0, b=1;
    int r=0;
    int max_iter = 10;
    while( max_iter-- )
    {
        r += (i++ == in) * a;
        std::swap(a,b);
        a += b;
    }
    return r;
}
\end{lstlisting}
\end{minipage}
\vspace{-0.2in}
\end{figure}

Changing a program to its data-oblivious form is not trivial. This unavoidable requirement reflects the fundamental property of computation where data is never decrypted, which is the case for all FHE computation. This constraint can affect usability, as existing algorithms may need modifications to be converted to a data-oblivious version.\footnote{Computer-assisted program transformation to data-oblivious variants is an active topic of research, focusing currently on domain-specific languages and compilers \cite{alchemy,fact}.} 
Moreover, it affects practicality too, as fixed iteration upper bounds, introduce performance degradation.

\subsection{FHE Batching}
The FHE batching technique offers the ability to pack several plaintexts into ``slots'' within a single ciphertext. This feature is supported by some FHE schemes, and in practice  enables parallel processing of plaintexts, which are part of the same ciphertext (in a Single Instruction Multiple Data (SIMD) style)~\cite{batching}. Notably, such SIMD processing enables significant performance improvements for algorithms with parallel computation properties.
Each ciphertext variable is effectively a vector of values and any unary or binary operation has the effect of array operations on all elements of these vectors. Using batching, integral data types naturally have arrays of bit values inside each bit of the variable, and gate operations on each separate bit have the effect of the gate operation on all bit values.
BGV, BFV, and CKKS all support batching which enables parallel processing of encrypted vectors of data.


\section{Proposed Bridging Methodology}\label{s:bridging}

We define new data types to abstract the low-level complexity of the encryption scheme and Boolean circuits.
This is possible by extending the  E3 framework \cite{e3eprint}, which is one level of abstraction above FHE libraries and enables portability of code across different libraries.
\subsection{Modular and bit-level arithmetic}\label{ss:modcom}

Some FHE schemes define arithmetic addition, subtraction, and multiplication on numeric rings, but not other arithmetic such as division or comparison. While some applications may not require any other operation, a programmer is accustomed to having all programming operations available in their programs.
For example, the C++ statement "{\tt{}if(a<b)c+=a}" and its corresponding data-oblivious form: "{\tt{}c+=(a<b)*a}", cannot be evaluated with an FHE scheme where the comparison operation is not defined.
Notably, addition, subtraction, and multiplication on integers is an incomplete set of arithmetic operations; for instance, the comparison operation cannot be reduced to these three operations. However, when operating on bits, the same set of operations is universal, having addition and subtraction correspond to logical \texttt{XOR} and multiplication to logical \texttt{AND}.

\eee\ solves this problem by allowing the programmer to use data types constructed out of sequences of encrypted bits. Indeed, as the least possible requirement to the computing capacity, an FHE scheme must be able to evaluate the logic \texttt{NAND} (or \texttt{NOR}) gate on ciphertexts, since this elementary function is sufficient for universal computation. 
In this way, the variables in the expression "{\tt{}c+=(a<b)*a}" are defined as \emph{integral types} where all three operations (comparison, multiplication, and addition) are performed using bit-level arithmetic circuits. We call this computation {\it bit-level arithmetic}, as opposed to the natively provided {\it modular arithmetic}, where addition and multiplication are performed directly on ciphertexts.

The transition from modular to bit-level arithmetic is straightforward: Since
the encrypted bit values are limited to 0 and~1, logic gates, such as \texttt{AND}, \texttt{XOR}, \texttt{NOT}, etc, can be expressed via the following expressions:
\vspace{-0.2cm}
\begin{equation*}\label{eq:logic}
\begin{split}
x \ \texttt{AND} \ y = xy,& \qquad x \ \texttt{NAND} \ y = 1-xy \\
x \ \texttt{OR} \ y = x+y-xy,& \qquad x \ \texttt{NOR} \ y = 1-(x \ \texttt{OR} \ y) \\
x \ \texttt{XOR} \ y = x+y-2xy,& \qquad x \ \texttt{XNOR} \ y = 1-(x \ \texttt{XOR} \ y) \\
 \texttt{NOT} \ x = 1-x,& \qquad \texttt{MUX}(x,y,z) = x(y-z)+z, \\
\end{split}
\end{equation*}

\noindent where \texttt{MUX} is the multiplexer operation (as in {$x$?$y$:$z$}). The set of values $\{0,1\}$ is closed under the above set of expressions. 
In this paper, we use the term \emph{homomorphic gates} to describe logic gates operating on encrypted bits. Given these homomorphic gates, higher level operations (comparison, addition, multiplication) can be built using standard combinational arithmetic circuit design, which allows to perform any programming operation, i.e. giving the full spectrum of C++ arithmetic. Gate equations simplify for plaintext modulo $2$; e.g. XOR gate does not require multiplication, as \mbox{$x \ \texttt{XOR} \ y = x+y \bmod 2$}.

\innersection{Data types}
We define three data types for bit-level arithmetic, \secuint, \secint, and \secbool, and one for modular arithmetic, \secmod. 
A \secuint\ or \secint\ variable is an array of ciphertexts, each encrypting either zero or one. \secuint\ is comparable to \texttt{unsigned int}, while \secint\ corresponds to \texttt{int}.
The third type is \secbool, which is a type derived from \secuint\texttt{<1>}. It is the secure equivalent of \texttt{bool}.
For modular arithmetic, there is only the native type, which operates modulo the plaintext modulus. We call this type \secmod.

With bit-level arithmetic, we can express programs with operations not supported by modular arithmetic.
Listing~\ref{list:fibs} demonstrates the transition from plaintext computation to secure computation: Integer variables are declared with the secure type \secuint; the rest of the code remains the same.
In encrypted computation, increasing the number of bits of integral types can lead to a dramatic increase in performance overhead.
For this reason, we define the bit size as a template specialization.
Explicit size declaration can be found in functional programming languages, so the programmer may already be familiar with this practice.
In the example of Listing~\ref{list:fibs}, all variables are 8 bits (lines 3, 4).

The postfix {\tt{}\_E} after each constant provides encrypted values used in variable initialization of integral type variables.
\eee\ automatically encrypts and replaces the plaintext constants with their encrypted representations, so the final binary does not have information about the initial values used in the program. 

\begin{figure}
\begin{minipage}{\linewidth}
\begin{lstlisting}[language=C++, caption={Secure version of Fibonacci function. Type \secuint\ behaves as native {\tt unsigned int}.
% \hspace{-1.40cm} \mbox{}
}, style=mystyle, label=list:fibs,
% xrightmargin=-0.12\linewidth,
% linewidth=0.96\linewidth
]
SecureUint<8> fibonacci(SecureUint<8> in)
{
    SecureUint<8> i=_0_E, a=_0_E, b=_1_E;
    SecureUint<8> r=_0_E;
    int max_iter = 10;
    while( max_iter-- )
    {
        r += (i++ == in) * a;
        std::swap(a,b);
        a += b;
    }
    return r;
}
\end{lstlisting}

\vspace{-0.5cm} 


\end{minipage}
\end{figure}

\subsection{Bridging modular and bit-level arithmetic}\label{ss:bridging}

Declaring variables with a protected integral type and using solely bit-level arithmetic as in Listing~\ref{list:fibs} has a potential drawback: When an FHE scheme provides fast modular arithmetic operations, the usage of circuits operating on separate bits is slow. This paper brings the  idea of 
\textit{Bridging} - mixing both modular and bit-level arithmetic in one program with the ability to convert variables from one type to the other. 
Some variables can be declared using a protected type supporting only modular arithmetic, while others with another secure type supporting bit-level arithmetic. In bridging mode, a type of bit-level arithmetic declares a conversion function into a type of modular arithmetic, and vice-versa.
In all cases, the encryption of the two different C++ types must share the same FHE keys, which is ensured by our proposed methodology and framework. 

\subsubsection{\secuint\ to \secmod{}}\label{sss:secuint2secmod}

For performance reasons, it is desirable to execute the entire computation in modular arithmetic, since it is much faster than bit-level. If however, a program requires an operation not supported in modular arithmetic (e.g., comparison), then, without Bridging, the whole program must perform all computations using bit-level arithmetic, severely degrading performance. 
In effect, Bridging enables the isolation of the parts of the computation requiring bit-level arithmetic.
For example, the expression "{\tt{}c+=(a<b)*a}" can use bit-level arithmetic for the comparison only.
The variables required by the comparison (i.e., \texttt{a} and \texttt{b}) must be of integral type. Nevertheless, the operands of the multiplication can be cast to our modular type, allowing multiplication and addition to be executed in modular arithmetic, resulting in a variable {\tt{}c} of modular type.

\begin{figure}
\begin{minipage}{\linewidth}
\begin{lstlisting}[language=C++, caption={
Bridging (i.e.,~mixing \secmod{} and \secuint\ types) enables performance improvement. The postfix \texttt{\_M} denotes encrypted variable for the \secmod\ type.
},
style=mystyle, 
label=list:fibm,
xleftmargin=0.45cm,
% xrightmargin=-0.12\linewidth,
% linewidth=0.88\linewidth
]
SecureMod fibonacci(SecureUint<8> in)
{
    SecureUint<8> i=_0_E;
    SecureMod a=_0_M, b=_1_M, r=_0_M;
    int max_iter = 10;
    while( max_iter-- )
    {
        r += (i++ == in) * a;
        std::swap(a,b);
        a += b;
    }
    return r;
}
\end{lstlisting}
\end{minipage}
\vspace{-0.8cm}

\end{figure}

Listing~\ref{list:fibm} demonstrates the code of the Fibonacci function of Listing~\ref{list:fibs} with \emph{Bridged} arithmetic: Only the input \texttt{in} and counter \texttt{i} are declared as integral type \secuint{}, while the others are replaced with the faster \secmod\ type. 
Line 8 does implicit conversion from \secuint\ to \secmod;
in this way, bit-level multiplication (which is slow) is not executed. Instead, only the native (much faster) multiplication of ciphertexts is needed.
Specifically, the comparison between {\tt i} and {\tt in} is the slowest operation in the program, and its result is one encrypted bit which can naturally be casted to \secmod, since the set $\{0,1\}$ is a subset of the plaintext range.
The operation for casting an integral type (bit-level) into modular (FHE native) is a summation of the encrypted bits of the integral type. 
In fact, the binary representation of a value $X$ can be reorganized by {\it Horner's scheme} in a set of additions over its $s$ bits $x_i$:
\begin{equation*}
\begin{split}
X & = 2^{s-1}x_{s-1} +2^{s-2}x_{s-2} + ... + 2x_1+x_0 = \\
& =(...((x_{s-1})\cdot 2+x_{s-2})\cdot 2+...+x_1)\cdot 2+x_0
\end{split}
\end{equation*}

Evaluating the right hand side of the above equation yields the value corresponding to the bit sequence. This evaluation is an efficient way to convert a program variable of type \secuint{} into a \secmod{} value.
Listing~\ref{list:secuint2secmod} shows the C++ implementation of such casting.
We emphasize that this conversion requires no ciphertext multiplications, only additions. Specifically, $2 \cdot (s - 1)$ ciphertext additions with a maximum additive depth of $s-1$, where $s$ is the number of encrypted bits.

\begin{figure}[t]
\begin{minipage}{\linewidth}
\begin{lstlisting}[language=C++, caption={
Casting from \secuint\ to \secmod. Since the former is a set of ciphertexts representing encrypted bits, it is possible to access each bit individually.}, style=mystyleb, label=list:secuint2secmod,
%framexrightmargin=-37pt,
% xleftmargin=0.4cm,
% xrightmargin=-0.12\linewidth,
% linewidth=0.88\linewidth
]
template <int Size>
SecureMod to_SecureMod(SecureUint<Size> v)
{
    auto i = Size;
    SecureMod r = v[--i];
    while ( i-- ) r += r + v[i];
    return r;
}
\end{lstlisting}
\end{minipage}
\vspace{-0.5cm} 
\end{figure}

\begin{figure}[t]
\begin{minipage}{\linewidth}
\begin{lstlisting}[language=C++, caption={
Casting from \secint\ to \secmod. The \secint\ variable is converted to \secuint, which is then converted to \secmod.}, style=mystyleb, label=list:secint2secmod,
%framexrightmargin=-37pt,
% xleftmargin=0.4cm,
% xrightmargin=-0.12\linewidth,
% linewidth=0.88\linewidth
]
template <int Size>
SecureMod to_SecureMod(SecureInt<Size> v)
{
    SecureUint<Size> u(v);
    auto pos = to_SecureMod(u);
    int max = 1 << Size;
    auto neg = SecureMod::t - max + pos;
    SecureMod isNeg = v[Size-1];
    return isNeg * neg + (1-isNeg) * pos;
}
\end{lstlisting}
\end{minipage}
\vspace{-0.7cm} 
\end{figure}


Line 8 of Listing~\ref{list:fibm} does implicit conversion of \secbool{} to \secmod{}. Note that \secbool\ is a derived class from \secuint\texttt{<1>}. To observe a more complex scenario, consider the expression "{\tt{}c+=(a==b)*a}" that actually requires conversion of \secuint{}.
Comparison between \texttt{a} and \texttt{b} must be evaluated in bit-level. This implies that types of \texttt{a} and \texttt{b} must be \secuint{}. The comparison is done on a bit-by-bit manner using homomorphic gates following the gate equations described in Section \ref{ss:modcom}.
The gates correspond to normal logic gates, but operating on ciphertexts instead of ordinary bits. The result of the comparison is one encrypted bit represented by type \secbool.
Multiplication between a \secbool\ and a \secuint\ is evaluated as a multiplexer operation with $s$ \texttt{AND} gates, where $s$ is the number of encrypted bits in the \secuint\ variable, resulting in a \secuint{} type.
The type of variable \texttt{c} can be chosen as \secmod{}. The addition in the expression is then performed on variables of \secmod{} and \secuint{} types: "{\tt{}c=c+t}", where "{\tt{}t=(a==b)*a}". Implicit conversion evokes our function of Listing \ref{list:secuint2secmod} from the constructor of \secmod{} type out of \secuint{}.
Then the addition operation follows on two variables, both of the \secmod{} type.

It should be noted that all these conversions and evaluations are done obliviously to the user and do not require special attention; the user writes only "{\tt{}c+=(a==b)*a}".
A more efficient way to perform this computation is to explicitly convert argument \texttt{a}  in this expression to \secmod. In such case, the multiplication is done in modular arithmetic which is around $s$ times faster than a multiplication between \secbool{} and \secuint{} (and many more times faster than multiplying two \secuint{}s).\footnote{The exact speed-up compared to multiplying two \secuint s depends on the variables' bit size.} The corresponding expression becomes \mbox{"{\tt{}c+=(a==b)*\secmod(a)}"}.
Same way as above, the constructor calls the conversion function (Listing~\ref{list:secuint2secmod}), this time one step earlier - before the multiplication - resulting in having a larger portion of the computation in modular arithmetic, hence improving the performance.
It it worth noting that while there is automatic conversion from \secuint\ to \secmod, it is the programmer's task to define each variable's type and, in some cases, call conversion explicitly for better performance.

\subsubsection{\secint\ to \secmod{}}\label{sss:secint2secmod}

So far, we have discussed unsigned numbers. However, bit-level arithmetic also supports signed numbers following the two's complement arithmetic.
On the other hand, modular arithmetic only supports numbers in $\mathbb{Z}_t$, where $t$ is the plaintext modulus.
Nevertheless, it is possible to emulate negative numbers in modular arithmetic in the programmer's domain as in Cryptoleq \cite{cryptoleq}, where lower values are considered positive and large values are interpreted as negative numbers.
In this case, the conversion from a signed bit-level arithmetic type \secint\ to \secmod\ is defined as:
\begin{equation}
  X=\begin{cases}
    t - 2^s + \sum_{i=0}^{s-1}{(2^i \cdot x_i)}, & \text{if $x<0$}.\\
    \sum_{i=0}^{s-1}{(2^i \cdot x_i)}, & \text{otherwise}.
  \end{cases}
\end{equation}
where $s$ is the number of bits and $x_i$ is the bit of $x$ at position $i$. The condition $x < 0$ is determined by the most significant bit of $x$; thus, we can use it as a multiplexer between the two cases.
Listing \ref{list:secint2secmod} presents the algorithm for converting a \secint\ into a \secmod. First, the \secint\ is interpreted as a \secuint\ (line 4). In line 5, this value is converted into a \secmod\ using the algorithm of Listing \ref{list:secuint2secmod}.
Then in line 7, we perform the subtraction of plaintexts $t$ and \texttt{max} (i.e. $2^s$), and then do a ciphertext addition with the \secmod\ variable \texttt{pos}. At this point, we have generated two \secmod\ variables: \texttt{pos}, representing the value in case it is positive, and \texttt{neg} containing the value in case it is a negative number. This totals $2s - 1$ ciphertext additions with a additive depth of $s$.
Finally, we select between \texttt{neg} and \texttt{pos} using the most significant bit of the \secint\ input (lines 8-9). If the bit is one, it means it is a negative number; thus, we select \texttt{neg}; otherwise, we select \texttt{pos}.
For selection we need ciphertext multiplications.
The entire conversion from \secint\ to \secmod\ requires two ciphertext multiplications and $2s + 1$ ciphertext additions with a multiplicative depth equal to one.
By comparing Listings \ref{list:secuint2secmod} and \ref{list:secint2secmod}, we can see that converting signed numbers to \secmod\ is less efficient than converting unsigned numbers to \secmod. Furthermore, $t$ must be large enough ($t \geq 2^s$) to accommodate the converted value.

\subsubsection{\secmod\ to \secuint{}}\label{sss:secmod2secuint}
\vspace{-0.2cm}
Conversion from \secmod\ to \secuint\ is theoretically possible using the expression:
\vspace{-0.2cm}
\begin{equation}\label{eq:secmod2secuint}
    \vspace{-0.1cm}
    X = \sum_{i=1}^{t-1}(i \cdot \secbool(1 - (x-i)^{t-1}))
\end{equation}
where $x$ is the \secmod\ variable to be converted, $t$ is the plaintext modulus and is prime, $i$ is a \secuint\ counter, and $X$ is the resulting \secuint. Due to the properties of modular arithmetic and the parameters used in homomorphic encryption, the exponentiation to $t-1$ results in zero in case the base is zero, and one otherwise. When $i = x$, the expression in the summation results in $i$, while in all other cases it will result in zero, since $1 - (x-i)^{t-1} = 0 \ \forall \ i \ne x$.

An implementation of the fast exponentiation algorithm is presented in Listing \ref{list:pow}. The number of multiplications is given by $\floor{\log_2{e}} + \omega(e) - 1$ and the multiplicative depth is $\ceil{\log_2{e}}$, where $e$ is the exponent and $\omega(\cdot)$ is a function that calculates the Hamming weight. The conversion from \secmod\ to \secuint\ can exploit this property of the exponentiation to $t-1$ resulting in zero or one to create an equality function, where the equality function is given by $1 - (x-i)^{t-1}$. With this information, we can build a linear search to find the \secuint\ $i$ that is equal to the \secmod\ $x$ using the result of the equality as a selector. Listing \ref{list:secmod2secuint} presents the algorithm that performs this conversion.
The number of multiplications is given by $t \cdot (s + \floor{\log_2{(t-1)}} + \omega(t-1) - 1)$, while the multiplicative depth is equal to $\ceil{\log_2{(t-1)}} + 1$.
One can notice that this algorithm is only practical for small plaintext moduli $t$. Once $t$ becomes large, the linear search makes it impractical. Since \secmod\ requires a large $t$ for it to be useful, this conversion should not be used in practice and should be avoided, as later presented and discussed in Section~\ref{ss:conversion}.

\begin{figure}[t]
\begin{minipage}{\linewidth}
\begin{lstlisting}[language=C++, caption={
Homomorphic exponentiation function.}, style=mystyleb, label=list:pow,
%framexrightmargin=-37pt,
% xleftmargin=0.4cm,
% xrightmargin=-0.12\linewidth,
% linewidth=0.88\linewidth
]
SecureMod pow(SecureMod b, int e)
{
    if (e == 0) return SecureMod(1);
    if (e == 1) return b;
    auto r = pow(b * b, e >> 1);
    if (e & 1) r *= b;
    return r;
}
\end{lstlisting}
\end{minipage}
\vspace{-0.5cm} 
\end{figure}

\begin{figure}[!t]
\begin{minipage}{\linewidth}
\begin{lstlisting}[language=C++, caption={
Casting from \secmod\ to \secuint.}, style=mystyleb, label=list:secmod2secuint,
%framexrightmargin=-37pt,
% xleftmargin=0.4cm,
% xrightmargin=-0.12\linewidth,
% linewidth=0.88\linewidth
]
template <int Size>
SecureUint to_SecureUint<Size>(SecureMod x)
{
SecureMod one(1);
    auto & t = SecureMod::t;
    vector<SecureUint<Size>> v;
    for (int i = 1; i < t; i++)
    {
        auto si = SecureUint<Size>(i);
        auto eq = one - pow(x-i, t-1);
        v.push_back(si * SecureBool(eq));
    }
    return sum(v);
}
\end{lstlisting}
\end{minipage}
\vspace{-0.5cm} 
\end{figure}

\subsubsection{\secmod\ to \secint{}}\label{sss:secmod2secint}

Conversion from \secmod\ to \secint\ is possible by applying Eq. \ref{eq:secmod2secint} to the result of Eq. \ref{eq:secmod2secuint}:
\begin{equation}\label{eq:secmod2secint}
    Y = (1 - X_{s-1}) \cdot X + X_{s-1} \cdot (2^s - t + X)
\end{equation}
$X$ is the result of Eq. \ref{eq:secmod2secuint}, $Y$ is the \secint\ output, $s$ is the number of encrypted bits in $X$ and $Y$, $t$ is the plaintext modulus, and $2^s \geq t$.
Listing \ref{list:secmod2secint} shows the algorithm for this conversion. It leverages the conversion described in Listing \ref{list:secmod2secuint} and adjusts for the sign.
This algorithm requires $2 \cdot t \cdot (s + \floor{\log_2{(t-1)}} + \omega(t-1) - 1)+2$ multiplications with a multiplicative depth of $\ceil{\log_2{(t-1)}} + 2$.

\begin{figure}[!t]
\begin{minipage}{\linewidth}
\begin{lstlisting}[language=C++, caption={
Casting from \secmod\ to \secint.}, style=mystyleb, label=list:secmod2secint,
%framexrightmargin=-37pt,
% xleftmargin=0.4cm,
% xrightmargin=-0.12\linewidth,
% linewidth=0.88\linewidth
]
template <int Size>
SecureInt to_SecureInt<Size>(SecureMod x)
{
    auto u = to_SecureUint<Size>(x);
    SecureInt<Size> pos(u);
    auto max = 1 << s;
    auto diff = max - SecureMod::t;
    u = to_SecureUint<Size>(diff + x);
    SecureInt<Size> neg(u);
    auto & isNeg = pos[s-1];
    return isNeg * neg + (1-isNeg) * pos;
}
\end{lstlisting}
\end{minipage}
\vspace{-0.5cm} 

\end{figure}

\section{Experimental Results}\label{s:results}

\subsection{Overview and setup}\label{ss:setup}

We evaluate bridging using three sets of experiments:
\begin{enumerate}
    \item Conversion from \secuint\ and \secint\ to \secmod, and vice-versa, using different bit sizes.
    \item Performance comparison of bridging and bit-level arithmetic using synthetic benchmarks.
    \item Case study analysis using one real-world FHE application.
\end{enumerate}


The results presented in Section \ref{s:results} were collected using a single thread (Sections \ref{ss:conversion} and \ref{ss:benchmarks}) and 24 threads (Section \ref{sss:genotype}) of an Intel Xeon Silver 4214R CPU @ 2.40GHz with 24 cores and 1 TB of memory running on RHEL 7.9, with GCC 7.3.1. We use the E3 framework (commit \#9fb718f) with SEAL 3.3.2 \cite{seal} as underlying FHE library, and BFV as the encryption scheme. We set the encryption parameters as: polynomial degree $n = 2^{15}$, as it provides the largest noise budget for encrypted computation, and plaintext modulus $t = 2^{16}+1$, as this is the smallest $t$ that enables batching for the chosen $n$. While a smaller $t$ would provide less noisy homomorphic operations, virtually all practical FHE applications rely on efficient utilization of batching. Nevertheless, although $t=2$ does not enable batching on BFV, we also test it on the benchmarks (Section \ref{ss:benchmarks}) since some homomorphic gates are simpler in modulo 2. The remaining parameters are automatically defined by SEAL given the required security level of 128 bits.
\subsection{Conversion}\label{ss:conversion}

\begin{figure}[t]
	\centering
    \frame{\includegraphics[width=\linewidth]{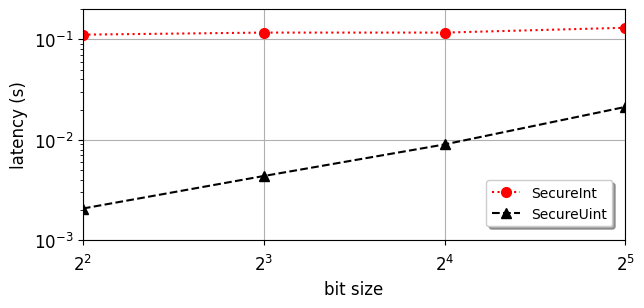}}
	\caption{Conversion latency time from \secuint\ and \secint\ to \secmod\ for different bit sizes. Throughput is up to $n=2^{15}$ times faster since a ciphertext fits $n$ plaintexts.}
	\label{fig:tosecmod}
	\vspace{-0.6cm} 
\end{figure}

Fig. \ref{fig:tosecmod} presents the conversion latency time from \secuint\ and \secint\ to \secmod\ for different bit sizes following the algorithms of Listings \ref{list:secuint2secmod} and \ref{list:secint2secmod}. The number of multiplications for \secuint\ and \secint\ stays constant at 0 and 2, while the multiplicative depth is 0 and 1, respectively, for all bit sizes. 
. As expected, conversion from both \secuint\ and \secint\ becomes slower for larger bit sizes due to the larger number of additions required. The slowdown is the same for both types. It is less noticeable for \secint\ due the log-scale graph and the fact that the multiplications dominate the \secint\ latency time.

\begin{figure}[t]
	\centering
    \frame{\includegraphics[width=\linewidth]{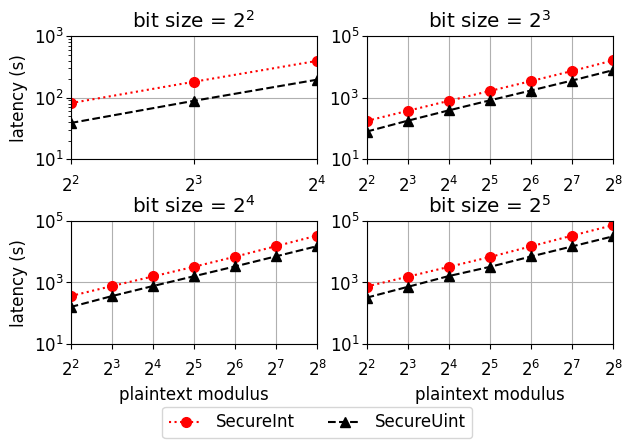}}
	\caption{Conversion latency time from \secmod\ to \secuint\ and \secint\ for different plaintext moduli and bit sizes.}
	\label{fig:fromsecmod}
	\vspace{-0.5cm}
\end{figure}

We also evaluate the conversion latency time from \secmod\ to \secuint/\secint\ using the algorithms presented in Listings \ref{list:secmod2secuint} and \ref{list:secmod2secint} with different plaintext moduli and bit sizes ($s = \{4, 8, 16, 32\}$). Fig.~\ref{fig:fromsecmod} summarizes the findings.
The results are what we expect from analysing the algorithms in Listings \ref{list:secmod2secuint} and \ref{list:secmod2secint}: When converting from \secmod\ to \secuint, the latency increases linearly to the bit size due to the operation on line 11 of Listing \ref{list:secmod2secuint}. Nevertheless, the dominant factor is the plaintext modulus $t$. A minor logarithmic effect comes from the exponentiation function (line 10) where the number of multiplications increases, while a major linear effect comes from the larger number of iterations in the \texttt{for} loop (line 7).
The conversion from \secmod\ to \secint\ (Listing \ref{list:secmod2secint}) takes roughly twice the time since it consists of two calls to the \secmod\ to \secuint\ conversion (lines 4 and 8) plus two multiplications (line 11).

\subsection{Benchmarks}\label{ss:benchmarks}

In order to evaluate different execution modes (bit-level arithmetic and bridging), we use six data-oblivious benchmarks, some of which were adapted from the TERMinator Suite \cite{terminator}. These benchmarks are developed to be data-oblivious and manipulate sets of encrypted variables adjusted to $\{4, 8, 16\}$-bit size.
The algorithms are: (FIB) Fibonacci, an additive-intensive algorithm, (LOG) Logistic Regression, where the data must be capped before inference, (MAX) Maximum, commonly used as non-linear function in ML applications, (MUX) Multiplexer, a simple operation similar to the ternary operator that replaces branch conditions on encrypted data, (PKS) Private Keyword Search, which searches privately for an item in a list or database, and (SOR) Sort, a sorting algorithm with low multiplicative depth, designed for FHE.


\begin{figure}[t]
	\centering
    \frame{\includegraphics[width=\linewidth]{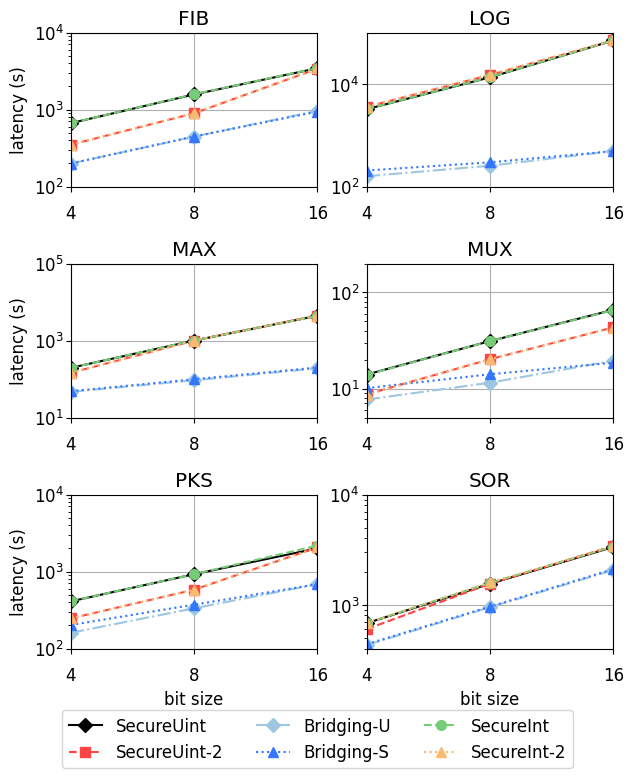}}
	\caption{Latency time for six benchmarks\iffalse using \secuint, \secint, unsigned and signed bridging, \texttt{Bridging-U} and \texttt{Bridging-S}, respectively,\fi with $t=2^{16}+1$. \iffalse In addition, we evaluate the benchmarks using \secuint\ and \secint\ with $t=2$ (postfix \texttt{-2}). \fi}
	\label{fig:benchmarks}
	\vspace{-0.4cm} 
\end{figure}

\innersection{Effect of plaintext modulus} Although $t=2$ does not enable batching in BFV, and therefore, it would not be used in practice, we compare it against the smallest plaintext modulus that enables batching ($t=2^{16}+1$) for $n=2^{15}$. The plaintext modulus does not affect the latency time to execute a homomorphic operation, but some homomorphic gates simplify in modulo 2 (e.g. XOR), as we discuss in Section \ref{ss:modcom}. Thus, a Boolean circuit used in bit-level arithmetic should have lower latency for $t=2$ if it contains XOR or XNOR gates. In Fig. \ref{fig:benchmarks}, we can see that in fact some applications are faster in modulo 2 (\texttt{SecureUint-2} and \texttt{SecureInt-2} for unsigned and signed bit-level arithmetic with $t=2$). MUX is the simplest benchmark, containing in bit-level arithmetic one equality, two \secbool-\secuint\ multiplications (or \secbool-\secint\ for signed numbers), one negation, and one addition. A \secbool-\secuint\ multiplication is entirely composed of XOR gates, and around half of the gates in the equality are XOR or XNOR. These operations are mainly responsible for the speed-up. 


\innersection{Bit-level arithmetic vs bridging} Fig. \ref{fig:benchmarks} presents the latency time comparison between bit-level arithmetic and bridging. Regarding unsigned numbers (\secuint\ vs \texttt{Bridging-U}), results show that bridging outperforms bit-level arithmetic for all benchmarks and bit sizes. For some applications, like SOR, the performance improvement is limited (around 60\% speed-up). This happens because this algorithm requires many non-native operations (comparisons) in most stages. Only in the last part of the algorithm bridging can be employed, since using bridging before would require the inefficient conversion from \secmod\ to \secuint\ in order to execute the latter comparisons. Conversely, the logistic regression benefits a lot from using bridging with a speed-up of more than two orders of magnitude. This is possible because the non-native operations required by the filtering function are executed first. Therefore, at the filtering stage it is already possible to employ bridging and perform the remaining computation using the faster modular arithmetic.

\innersection{Signed numbers} Also in Fig. \ref{fig:benchmarks} we compare how bridging behaves with signed numbers. We can see some degradation in PKS (4 bits) and MUX (4 and 8 bits). In PKS, there is conversion from \secint\ to \secmod\ in every iteration of the loop. The slowest operation in the loop is the comparison. This comparison operation is faster for smaller circuits (4 bits); therefore, the proportional latency of the two ciphertext multiplications required for the \secint\ to \secmod\ conversion is higher. For larger circuits, the comparison becomes more costly, thus, amortizing the conversion cost.


\begin{figure}[t]
	\centering
    \frame{\includegraphics[width=\linewidth]{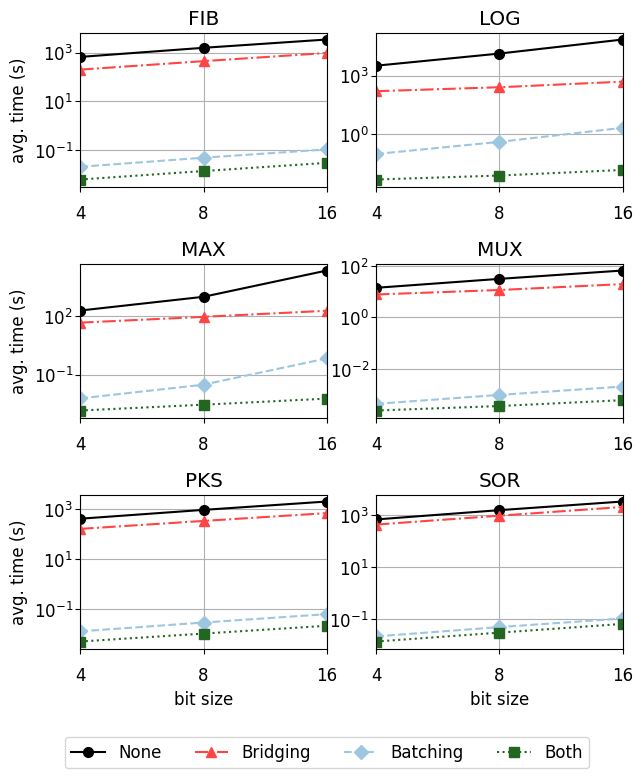}}
	\caption{Average execution time for six benchmarks. \iffalse using and four cases: \texttt{None}: without bridging and batching, \texttt{Bridging}: with bridging only, \texttt{Batching}: with batching only, and \texttt{Both}: with both bridging and batching.\fi}
	\label{fig:batching}
	\vspace{-0.4cm} 
\end{figure}

\innersection{Bridging can be used with batching} Batching is a powerful technique used in FHE to pack many plaintexts into a ciphertext. The number of plaintexts that can be packed into a ciphertext is equal to the polynomial degree ($n = 2^{15}$ in our case). Although carefully crafted algorithms using batching and rotations can reduce program latency, the main benefit of batching is increased throughput, since it is possible to process $n$ plaintexts at once. Bridging on the other hand always reduces latency, and consequently improves throughput.
Bridging is a technique independent of batching and both can be used in tandem.
In Fig. \ref{fig:batching} we present the average execution time ($\text{latency} / \text{\# plaintexts packed}$) for six benchmarks and four cases: no batching and no bridging (\texttt{None}), bridging-only (\texttt{Bridging}), batching-only (\texttt{Batching}), and batching and bridging together (\texttt{Both}).
As expected, the throughput improvement provided by batching is impressive. However, for this technique to be used at its fullest, it is necessary to fill all ciphertext slots with plaintexts, which is the case with embarrassingly parallel workload. If the load is less than $n$, the performance will degrade.
Bridging does not suffer from this problem since it focuses on reducing the latency.
Nevertheless, bridging can be used in combination with batching. In this case, the speed-up of both techniques is added, leading to an even faster runtime (acceleration by more than 6 orders of magnitude (LOG)).

\innersection{Insights}
Bridging demonstrates substantial performance improvement for algorithms that allow mixing \secuint{} and \secmod{}, while in the worst case, bridging is equivalent to bit-level arithmetic. 
The comparison heavy SOR benefits less from bridging since these operations occur throughout the program and near the output, which precludes using the \secmod{} type in an earlier stage. On the contrary the benchmarks containing non-native operations near the beginning exhibit more significant performance improvements (MAX, FIB, PKS).
Bridging benefits are maximized when non-native operations (e.g. comparisons) are at the beginning of the computation and the remaining of the computation can be done in modular arithmetic. The non-native operation would require the whole computation to work on bit-level arithmetic, but with bridging it becomes much faster since only the non-native operations are performed with \secuint, while the remaining run using \secmod. This is demonstrated by our logistic regression benchmark, which provides more than two orders of magnitude (precisely 143 times) of performance improvement.


\vspace{-0.1cm}

\subsection{Case-study: Genotype imputation}\label{sss:genotype}

Genotype imputation is the process of filling missing information in DNA sequencing with the use of statistical methods. P-Impute \cite{GURSOY2021} converts an inefficient statistical model based on correlation of similar individuals 
into a Private Information Retrieval (PIR) problem, which has efficient solutions in FHE, using the BFV encryption scheme. In this section, we discuss the performance of p-Impute without (\secuint) and with bridging taking into consideration the multiplicative depth for defining more efficient encryption parameters. Batching is used to fit $n$ plaintexts into a ciphertext.

We set each run to use 24 threads and report the query time, i.e., the time performing encrypted computation, and the number of ciphertext additions, multiplications, and subtractions in Table \ref{tab:nops}. We use the same plaintext modulus as in the other experiments ($t = 2^{16}+1$), but we vary the polynomial degree ($n \in \{2^{13}, 2^{14}, 2^{15}\}$) in order to evaluate the extra performance improvement provided by bridging for requiring a lower multiplicative depth.
Without bridging, we were only able to run with $n = 2^{15}$, since lower polynomial degrees do not provide enough noise budget to run the application without bootstrapping. 
With bridging, we were able to reduce $n$ to $2^{13}$ without corrupting the result.
As $n$ reduces, the number of batching slots in a ciphertext reduces since it is equal to the polynomial degree. However, the reduction in latency for the homomorphic operation compensates for the reduction in slots. Experimental results show that when $n$ is halved, the latency of the ciphertext multiplication reduces by at least 4 times, and can be amplified depending on the behavior of cache memories. Consequently, one halving of $n$ increases throughput by at least 2x and reduces latency by 4x.

For the same polynomial degree ($n = 2^{15}$), bridging is around 6.82x faster than bit-level arithmetic. This is due to the reduced number of operations on encrypted data, as one can see in Table \ref{tab:nops}.
Nevertheless, bridging requires less noise budget to operate, which allows us to use smaller polynomial degrees. In the fastest case, bridging has the throughput improved by 62.2x, while the latency reduces by around 249 times.
\vspace{-0.1cm}
\begin{table}[t]
    \centering
    \caption{Number of ciphertext \iffalse additions, multiplications, and subtractions, \fi operations and execution time for p-Impute \cite{GURSOY2021}. \iffalse with and without bridging \replace[and]{using} different polynomial degrees (n).  Without bridging, only $n = 2^{15}$ has enough noise budget for the computation. \fi}
    \vspace{-0.4cm}
    \begin{tabular}{crrrrr}
        Bridging & n &  add  &  mul  &  sub  & time (s) \\ \hline
        \normalfont{no}  & \unboldmath \( 2^{15} \) & \normalfont{53288} & \normalfont{66152} & \normalfont{51256} & \normalfont{5026} \\ \hline
        \normalfont{yes} & \unboldmath \( 2^{15} \) & \normalfont{14248} & \normalfont{ 9072} & \normalfont{ 4536} & \normalfont{ 737} \\ \hline
        \normalfont{no} & \unboldmath \( 2^{14} \) & \normalfont{ NA} & \normalfont{ NA} & \normalfont{ NA} & \normalfont{ NA} \\ 
        \hline
        \normalfont{yes} & \unboldmath \( 2^{14} \) & \normalfont{28496} & \normalfont{18144} & \normalfont{ 9072} & \normalfont{ 196} \\ 
        \hline  
        \normalfont{no} & \unboldmath \( 2^{13} \) & \normalfont{ NA} & \normalfont{ NA} & \normalfont{ NA} & \normalfont{ NA} \\ 
        \hline
        \normalfont{yes} & \unboldmath \( 2^{13} \) & \normalfont{56992} & \normalfont{36288} & \normalfont{18144} & \normalfont{  80} \\ 
             
    \end{tabular}
    \label{tab:nops}
    \vspace{-0.5cm}
\end{table}

\section{Discussion}\label{s:related_work}

\noindent {\bf Polynomial degree:}
We compare bridging with other accelerations of homomorphic computation at the programming level.
As we show in Section \ref{sss:genotype}, the polynomial degree affects the performance and noise budget. A smaller polynomial degree makes operations a few times faster. At the same time, it reduces the noise budget, allowing fewer computations before data corruption or bootstrapping. It also reduces the number of plaintexts that can be packed in a ciphertext when batching. Nevertheless, the performance improvement from a smaller polynomial degree outweighs the fewer batching slots in the ciphertexts.

\noindent {\bf Batching} enables SIMD usage of ciphertexts \cite{batching}. It can provide several orders of magnitude performance improvement for SIMD-compatible applications, however not all FHE schemes support it \cite{chillotti2016faster,ducas2015fhew}. Due to its significant computational benefits, batching should always be used where possible. 


\noindent {\bf Bridging} makes possible to use the comprehensive bit-level arithmetic and the fast modular arithmetic in the same program. It increases the expressivity of programs previously limited to modular arithmetic, and improves performance of complex programs implemented using bit-level arithmetic. At the same time, it reduces noise, since the depth of the datapath is shorter due to fewer homomorphic operations being executed, which may enable using a smaller polynomial degree, further improving performance.
Therefore, the performance improvement provided by bridging comes from two factors: 1) Reduced number of homomorphic operations since Boolean circuits performing additions and multiplications in bit-level arithmetic are replaced by a single operation (native addition or multiplication), and 2) reduced multiplicative depth, since when Boolean circuits are replaced by a single instruction, the multiplicative depth reduces. This reduces the noise budget required by the application, enabling the use of a smaller polynomial degree, further improving performance.
In addition, bridging is independent of encryption parameters, apart from the plaintext modulus which should not be too small. This makes bridging perfectly compatible with batching and any polynomial degree.

\section{Conclusions}\label{s:conclusions}

In this work, we presented a new methodology combining universal bit-level arithmetic and faster modular arithmetic computation, dubbed \emph{bridging}, to accelerate applications using fully homomorphic encryption. Experiments demonstrate significant performance improvements when using both bridging and batching modes. Bridging by itself can offer several orders of magnitude performance improvement, depending on the type of application. In the benchmark evaluation, bridging improved performance by more than 2 orders, and 6 orders of magnitude when combined with batching. 
Furthermore, the case-study private genotype imputation became two orders of magnitude faster due to reduced number of homomorphic operations and multiplicative depth, allowing us to use more efficient encryption parameters.

\section*{Resources}

Bridging has been integrated with the E3 framework and is available on github.com/momalab/e3 from commit \#88a323f.

\bibliographystyle{ACM-Reference-Format}
\bibliography{reference}


\begin{thebibliography}{33}


\ifx \showCODEN    \undefined \def \showCODEN     #1{\unskip}     \fi
\ifx \showDOI      \undefined \def \showDOI       #1{#1}\fi
\ifx \showISBNx    \undefined \def \showISBNx     #1{\unskip}     \fi
\ifx \showISBNxiii \undefined \def \showISBNxiii  #1{\unskip}     \fi
\ifx \showISSN     \undefined \def \showISSN      #1{\unskip}     \fi
\ifx \showLCCN     \undefined \def \showLCCN      #1{\unskip}     \fi
\ifx \shownote     \undefined \def \shownote      #1{#1}          \fi
\ifx \showarticletitle \undefined \def \showarticletitle #1{#1}   \fi
\ifx \showURL      \undefined \def \showURL       {\relax}        \fi
\providecommand\bibfield[2]{#2}
\providecommand\bibinfo[2]{#2}
\providecommand\natexlab[1]{#1}
\providecommand\showeprint[2][]{arXiv:#2}

\bibitem[Adee(2008)]%
        {adee2008hunt}
\bibfield{author}{\bibinfo{person}{Sally Adee}.}
  \bibinfo{year}{2008}\natexlab{}.
\newblock \showarticletitle{The hunt for the kill switch}.
\newblock \bibinfo{journal}{\emph{IEEE Spectrum}} \bibinfo{volume}{45},
  \bibinfo{number}{5} (\bibinfo{year}{2008}), \bibinfo{pages}{34--39}.
\newblock


\bibitem[Brakerski et~al\mbox{.}(2014)]%
        {BGV_ref}
\bibfield{author}{\bibinfo{person}{Zvika Brakerski}, \bibinfo{person}{Craig
  Gentry}, {and} \bibinfo{person}{Vinod Vaikuntanathan}.}
  \bibinfo{year}{2014}\natexlab{}.
\newblock \showarticletitle{(Leveled) fully homomorphic encryption without
  bootstrapping}.
\newblock \bibinfo{journal}{\emph{ACM Transactions on Computation Theory
  (TOCT)}} \bibinfo{volume}{6}, \bibinfo{number}{3} (\bibinfo{year}{2014}),
  \bibinfo{pages}{1--36}.
\newblock


\bibitem[Cauligi et~al\mbox{.}(2019)]%
        {fact}
\bibfield{author}{\bibinfo{person}{Sunjay Cauligi}, \bibinfo{person}{Gary
  Soeller}, \bibinfo{person}{Brian Johannesmeyer}, \bibinfo{person}{Fraser
  Brown}, \bibinfo{person}{Riad~S Wahby}, \bibinfo{person}{John Renner},
  \bibinfo{person}{Benjamin Gr{\'e}goire}, \bibinfo{person}{Gilles Barthe},
  \bibinfo{person}{Ranjit Jhala}, {and} \bibinfo{person}{Deian Stefan}.}
  \bibinfo{year}{2019}\natexlab{}.
\newblock \showarticletitle{FaCT: a DSL for timing-sensitive computation}. In
  \bibinfo{booktitle}{\emph{ACM Conference on Programming Language Design and
  Implementation}}. \bibinfo{pages}{174--189}.
\newblock


\bibitem[Chen et~al\mbox{.}(2018)]%
        {chen2018sgxpectre}
\bibfield{author}{\bibinfo{person}{Guoxing Chen}, \bibinfo{person}{Sanchuan
  Chen}, \bibinfo{person}{Yuan Xiao}, \bibinfo{person}{Yinqian Zhang},
  \bibinfo{person}{Zhiqiang Lin}, {and} \bibinfo{person}{Ten~H Lai}.}
  \bibinfo{year}{2018}\natexlab{}.
\newblock \showarticletitle{{SGXspectre Attacks: Leaking Enclave Secrets via
  Speculative Execution}}.
\newblock \bibinfo{journal}{\emph{arXiv preprint arXiv:1802.09085}}
  (\bibinfo{year}{2018}).
\newblock


\bibitem[Cheon et~al\mbox{.}(2017)]%
        {CKKS_ref}
\bibfield{author}{\bibinfo{person}{Jung~Hee Cheon}, \bibinfo{person}{Andrey
  Kim}, \bibinfo{person}{Miran Kim}, {and} \bibinfo{person}{Yongsoo Song}.}
  \bibinfo{year}{2017}\natexlab{}.
\newblock \showarticletitle{Homomorphic encryption for arithmetic of
  approximate numbers}. In \bibinfo{booktitle}{\emph{ASIACRYPT}}. Springer,
  \bibinfo{pages}{409--437}.
\newblock


\bibitem[Chielle et~al\mbox{.}(2018)]%
        {e3eprint}
\bibfield{author}{\bibinfo{person}{Eduardo Chielle}, \bibinfo{person}{Oleg
  Mazonka}, \bibinfo{person}{Nektarios~Georgios Tsoutsos}, {and}
  \bibinfo{person}{Michail Maniatakos}.} \bibinfo{year}{2018}\natexlab{}.
\newblock \bibinfo{title}{E$^3$: A Framework for Compiling C++ Programs with
  Encrypted Operands}.
\newblock \bibinfo{howpublished}{Cryptology ePrint Archive, Report 2018/1013}.
\newblock
\newblock
\shownote{Online: \url{https://eprint.iacr.org/2018/1013}, GitHub repository:
  \url{https://github.com/momalab/e3}}.


\bibitem[Chillotti et~al\mbox{.}(2016)]%
        {chillotti2016faster}
\bibfield{author}{\bibinfo{person}{Ilaria Chillotti}, \bibinfo{person}{Nicolas
  Gama}, \bibinfo{person}{Mariya Georgieva}, {and} \bibinfo{person}{Malika
  Izabachene}.} \bibinfo{year}{2016}\natexlab{}.
\newblock \showarticletitle{Faster fully homomorphic encryption: Bootstrapping
  in less than 0.1 seconds}. In \bibinfo{booktitle}{\emph{ASIACRYPT}}.
  Springer, \bibinfo{pages}{3--33}.
\newblock


\bibitem[Costan and Devadas(2016)]%
        {costan2016sgx-explained}
\bibfield{author}{\bibinfo{person}{Victor Costan} {and}
  \bibinfo{person}{Srinivas Devadas}.} \bibinfo{year}{2016}\natexlab{}.
\newblock \showarticletitle{{Intel SGX Explained}}.
\newblock \bibinfo{journal}{\emph{IACR Cryptology ePrint Archive}}
  \bibinfo{volume}{2016} (\bibinfo{year}{2016}), \bibinfo{pages}{086}.
\newblock


\bibitem[Crockett et~al\mbox{.}(2018)]%
        {alchemy}
\bibfield{author}{\bibinfo{person}{Eric Crockett}, \bibinfo{person}{Chris
  Peikert}, {and} \bibinfo{person}{Chad Sharp}.}
  \bibinfo{year}{2018}\natexlab{}.
\newblock \showarticletitle{Alchemy: A language and compiler for homomorphic
  encryption made easy}. In \bibinfo{booktitle}{\emph{ACM Conference on
  Computer and Communications Security (CCS)}}. \bibinfo{pages}{1020--1037}.
\newblock


\bibitem[Ducas and Micciancio(2015)]%
        {ducas2015fhew}
\bibfield{author}{\bibinfo{person}{L{\'e}o Ducas} {and}
  \bibinfo{person}{Daniele Micciancio}.} \bibinfo{year}{2015}\natexlab{}.
\newblock \showarticletitle{{FHEW}: bootstrapping homomorphic encryption in
  less than a second}. In \bibinfo{booktitle}{\emph{EUROCRYPT}}. Springer,
  \bibinfo{pages}{617--640}.
\newblock


\bibitem[Fan and Vercauteren(2012)]%
        {fan2012somewhatmisc}
\bibfield{author}{\bibinfo{person}{Junfeng Fan} {and} \bibinfo{person}{Frederik
  Vercauteren}.} \bibinfo{year}{2012}\natexlab{}.
\newblock \showarticletitle{Somewhat Practical Fully Homomorphic Encryption.}
\newblock \bibinfo{journal}{\emph{IACR Cryptology ePrint Archive}}
  \bibinfo{volume}{2012} (\bibinfo{year}{2012}), \bibinfo{pages}{144}.
\newblock


\bibitem[Gentry et~al\mbox{.}(2013)]%
        {GSW_ref}
\bibfield{author}{\bibinfo{person}{Craig Gentry}, \bibinfo{person}{Amit Sahai},
  {and} \bibinfo{person}{Brent Waters}.} \bibinfo{year}{2013}\natexlab{}.
\newblock \showarticletitle{Homomorphic encryption from learning with errors:
  Conceptually-simpler, asymptotically-faster, attribute-based}. In
  \bibinfo{booktitle}{\emph{CRYPTO}}. Springer, \bibinfo{pages}{75--92}.
\newblock


\bibitem[Gross(2018)]%
        {gross2018ending}
\bibfield{author}{\bibinfo{person}{Judah~Ari Gross}.}
  \bibinfo{year}{2018}\natexlab{}.
\newblock \bibinfo{title}{Ending a decade of silence, Israel confirms it blew
  up Assad's nuclear reactor}.
\newblock
  \bibinfo{howpublished}{\url{https://www.timesofisrael.com/ending-a-decade-of-silence-israel-reveals-it-blew-up-assads-nuclear-reactor/}}.
\newblock


\bibitem[Gürsoy et~al\mbox{.}(2021)]%
        {GURSOY2021}
\bibfield{author}{\bibinfo{person}{Gamze Gürsoy}, \bibinfo{person}{Eduardo
  Chielle}, \bibinfo{person}{Charlotte~M. Brannon}, \bibinfo{person}{Michail
  Maniatakos}, {and} \bibinfo{person}{Mark Gerstein}.}
  \bibinfo{year}{2021}\natexlab{}.
\newblock \showarticletitle{Privacy-preserving genotype imputation with fully
  homomorphic encryption}.
\newblock \bibinfo{journal}{\emph{Cell Systems}} (\bibinfo{year}{2021}).
\newblock
\showISSN{2405-4712}
\urldef\tempurl%
\url{https://doi.org/10.1016/j.cels.2021.10.003}
\showDOI{\tempurl}


\bibitem[Halevi and Shoup(2018)]%
        {cryptoeprint:2018:244}
\bibfield{author}{\bibinfo{person}{Shai Halevi} {and} \bibinfo{person}{Victor
  Shoup}.} \bibinfo{year}{2018}\natexlab{}.
\newblock \bibinfo{title}{Faster Homomorphic Linear Transformations in HElib}.
\newblock \bibinfo{howpublished}{Cryptology ePrint Archive, Report 2018/244}.
\newblock
\newblock
\shownote{\url{https://eprint.iacr.org/2018/244}}.


\bibitem[H{\'e}ly et~al\mbox{.}(2012)]%
        {hely2012malicious}
\bibfield{author}{\bibinfo{person}{David H{\'e}ly}, \bibinfo{person}{Maurin
  Augagneur}, \bibinfo{person}{Yves Clauzel}, {and}
  \bibinfo{person}{J{\'e}r{\'e}my Dubeuf}.} \bibinfo{year}{2012}\natexlab{}.
\newblock \showarticletitle{Malicious key emission via hardware Trojan against
  encryption system}. In \bibinfo{booktitle}{\emph{International Conference on
  Computer Design (ICCD)}}. IEEE, \bibinfo{pages}{127--130}.
\newblock


\bibitem[Ibarrondo and Viand(2021)]%
        {pyfhel}
\bibfield{author}{\bibinfo{person}{Alberto Ibarrondo} {and}
  \bibinfo{person}{Alexander Viand}.} \bibinfo{year}{2021}\natexlab{}.
\newblock \showarticletitle{Pyfhel: PYthon For Homomorphic Encryption
  Libraries}. In \bibinfo{booktitle}{\emph{Proceedings of the 9th on Workshop
  on Encrypted Computing; Applied Homomorphic Cryptography}} (Virtual Event,
  Republic of Korea) \emph{(\bibinfo{series}{WAHC '21})}.
  \bibinfo{publisher}{Association for Computing Machinery},
  \bibinfo{address}{New York, NY, USA}, \bibinfo{pages}{11–16}.
\newblock
\showISBNx{9781450386562}
\urldef\tempurl%
\url{https://doi.org/10.1145/3474366.3486923}
\showDOI{\tempurl}


\bibitem[Jin et~al\mbox{.}(2012)]%
        {jin2012exposing}
\bibfield{author}{\bibinfo{person}{Yier Jin}, \bibinfo{person}{Michail
  Maniatakos}, {and} \bibinfo{person}{Yiorgos Makris}.}
  \bibinfo{year}{2012}\natexlab{}.
\newblock \showarticletitle{Exposing vulnerabilities of untrusted computing
  platforms}. In \bibinfo{booktitle}{\emph{International Conference on Computer
  Design (ICCD)}}. IEEE, \bibinfo{pages}{131--134}.
\newblock


\bibitem[Karri et~al\mbox{.}(2010)]%
        {karri2010trustworthy}
\bibfield{author}{\bibinfo{person}{Ramesh Karri}, \bibinfo{person}{Jeyavijayan
  Rajendran}, \bibinfo{person}{Kurt Rosenfeld}, {and} \bibinfo{person}{Mohammad
  Tehranipoor}.} \bibinfo{year}{2010}\natexlab{}.
\newblock \showarticletitle{Trustworthy hardware: Identifying and classifying
  hardware trojans}.
\newblock \bibinfo{journal}{\emph{Computer}} \bibinfo{volume}{43},
  \bibinfo{number}{10} (\bibinfo{year}{2010}), \bibinfo{pages}{39--46}.
\newblock


\bibitem[Kocher et~al\mbox{.}(2019)]%
        {Kocher2018spectre}
\bibfield{author}{\bibinfo{person}{Paul Kocher}, \bibinfo{person}{Jann Horn},
  \bibinfo{person}{Anders Fogh}, \bibinfo{person}{}, \bibinfo{person}{Daniel
  Genkin}, \bibinfo{person}{Daniel Gruss}, \bibinfo{person}{Werner Haas},
  \bibinfo{person}{Mike Hamburg}, \bibinfo{person}{Moritz Lipp},
  \bibinfo{person}{Stefan Mangard}, \bibinfo{person}{Thomas Prescher},
  \bibinfo{person}{Michael Schwarz}, {and} \bibinfo{person}{Yuval Yarom}.}
  \bibinfo{year}{2019}\natexlab{}.
\newblock \showarticletitle{Spectre Attacks: Exploiting Speculative Execution}.
  In \bibinfo{booktitle}{\emph{40th IEEE Symposium on Security and Privacy
  (S\&P'19)}}.
\newblock


\bibitem[Lipp et~al\mbox{.}(2018)]%
        {Lipp2018meltdown}
\bibfield{author}{\bibinfo{person}{Moritz Lipp}, \bibinfo{person}{Michael
  Schwarz}, \bibinfo{person}{Daniel Gruss}, \bibinfo{person}{Thomas Prescher},
  \bibinfo{person}{Werner Haas}, \bibinfo{person}{Anders Fogh},
  \bibinfo{person}{Jann Horn}, \bibinfo{person}{Stefan Mangard},
  \bibinfo{person}{Paul Kocher}, \bibinfo{person}{Daniel Genkin},
  \bibinfo{person}{Yuval Yarom}, {and} \bibinfo{person}{Mike Hamburg}.}
  \bibinfo{year}{2018}\natexlab{}.
\newblock \showarticletitle{Meltdown: Reading Kernel Memory from User Space}.
  In \bibinfo{booktitle}{\emph{27th {USENIX} Security Symposium ({USENIX}
  Security 18)}}.
\newblock


\bibitem[Mazonka et~al\mbox{.}(2016)]%
        {cryptoleq}
\bibfield{author}{\bibinfo{person}{Oleg Mazonka},
  \bibinfo{person}{Nektarios~Georgios Tsoutsos}, {and} \bibinfo{person}{Michail
  Maniatakos}.} \bibinfo{year}{2016}\natexlab{}.
\newblock \showarticletitle{Cryptoleq: A Heterogeneous Abstract Machine for
  Encrypted and Unencrypted Computation}.
\newblock \bibinfo{journal}{\emph{IEEE Transactions on Information Forensics
  and Security}} \bibinfo{volume}{11}, \bibinfo{number}{9}
  (\bibinfo{year}{2016}), \bibinfo{pages}{2123--2138}.
\newblock
\urldef\tempurl%
\url{https://doi.org/10.1109/TIFS.2016.2569062}
\showDOI{\tempurl}


\bibitem[Mouchet et~al\mbox{.}(2020)]%
        {lattigop}
\bibfield{author}{\bibinfo{person}{Christian Mouchet},
  \bibinfo{person}{Jean-Philippe Bossuat}, \bibinfo{person}{Juan~Ram{\'o}n
  Troncoso-Pastoriza}, {and} \bibinfo{person}{Jean-Pierre Hubaux}.}
  \bibinfo{year}{2020}\natexlab{}.
\newblock \showarticletitle{Lattigo: a Multiparty Homomorphic Encryption
  Library in Go}.
\newblock


\bibitem[Mouris et~al\mbox{.}(2018)]%
        {terminator}
\bibfield{author}{\bibinfo{person}{Dimitris Mouris},
  \bibinfo{person}{Nektarios~Georgios Tsoutsos}, {and} \bibinfo{person}{Michail
  Maniatakos}.} \bibinfo{year}{2018}\natexlab{}.
\newblock \showarticletitle{TERMinator Suite: Benchmarking Privacy-Preserving
  Architectures}.
\newblock \bibinfo{journal}{\emph{IEEE Computer Architecture Letters}}
  \bibinfo{volume}{17}, \bibinfo{number}{2} (\bibinfo{year}{2018}),
  \bibinfo{pages}{122--125}.
\newblock
\showISSN{1556-6056}


\bibitem[nuFHE(2018)]%
        {nuFHE}
nuFHE \bibinfo{year}{2018}\natexlab{}.
\newblock \bibinfo{title}{{NuCypher fully homomorphic encryption (NuFHE)}}.
\newblock \bibinfo{howpublished}{\url{https://github.com/nucypher/nufhe}}.
\newblock
\newblock
\shownote{NuCypher}.


\bibitem[Pinto and Santos(2019)]%
        {trustzone}
\bibfield{author}{\bibinfo{person}{Sandro Pinto} {and} \bibinfo{person}{Nuno
  Santos}.} \bibinfo{year}{2019}\natexlab{}.
\newblock \showarticletitle{Demystifying Arm TrustZone: A Comprehensive
  Survey}.
\newblock \bibinfo{journal}{\emph{ACM Comput. Surv.}} \bibinfo{volume}{51},
  \bibinfo{number}{6}, Article \bibinfo{articleno}{130} (\bibinfo{date}{Jan.}
  \bibinfo{year}{2019}), \bibinfo{numpages}{36}~pages.
\newblock
\showISSN{0360-0300}
\urldef\tempurl%
\url{https://doi.org/10.1145/3291047}
\showDOI{\tempurl}


\bibitem[Samardzic et~al\mbox{.}(2021)]%
        {f1}
\bibfield{author}{\bibinfo{person}{Nikola Samardzic}, \bibinfo{person}{Axel
  Feldmann}, \bibinfo{person}{Aleksandar Krastev}, \bibinfo{person}{Srinivas
  Devadas}, \bibinfo{person}{Ronald Dreslinski}, \bibinfo{person}{Christopher
  Peikert}, {and} \bibinfo{person}{Daniel Sanchez}.}
  \bibinfo{year}{2021}\natexlab{}.
\newblock \showarticletitle{F1: A Fast and Programmable Accelerator for Fully
  Homomorphic Encryption}. In \bibinfo{booktitle}{\emph{MICRO-54: 54th Annual
  IEEE/ACM International Symposium on Microarchitecture}} (Virtual Event,
  Greece) \emph{(\bibinfo{series}{MICRO '21})}. \bibinfo{publisher}{Association
  for Computing Machinery}, \bibinfo{address}{New York, NY, USA},
  \bibinfo{pages}{238–252}.
\newblock
\showISBNx{9781450385572}
\urldef\tempurl%
\url{https://doi.org/10.1145/3466752.3480070}
\showDOI{\tempurl}


\bibitem[SEAL(2019)]%
        {seal}
SEAL \bibinfo{year}{2019}\natexlab{}.
\newblock \bibinfo{title}{{M}icrosoft {SEAL} (release 3.3.2)}.
\newblock \bibinfo{howpublished}{\url{https://github.com/Microsoft/SEAL}}.
\newblock
\newblock
\shownote{Microsoft Research, Redmond, WA.}.


\bibitem[Smart and Vercauteren(2014)]%
        {batching}
\bibfield{author}{\bibinfo{person}{Nigel~P. Smart} {and}
  \bibinfo{person}{Frederik Vercauteren}.} \bibinfo{year}{2014}\natexlab{}.
\newblock \showarticletitle{Fully homomorphic {SIMD} operations}.
\newblock \bibinfo{journal}{\emph{Designs, codes and cryptography}}
  \bibinfo{volume}{71}, \bibinfo{number}{1} (\bibinfo{year}{2014}),
  \bibinfo{pages}{57--81}.
\newblock


\bibitem[Tsoutsos and Maniatakos(2014)]%
        {tsoutsos2013fabrication}
\bibfield{author}{\bibinfo{person}{Nektarios~Georgios Tsoutsos} {and}
  \bibinfo{person}{Michail Maniatakos}.} \bibinfo{year}{2014}\natexlab{}.
\newblock \showarticletitle{Fabrication attacks: {Z}ero-overhead malicious
  modifications enabling modern microprocessor privilege escalation}.
\newblock \bibinfo{journal}{\emph{IEEE Transactions on Emerging Topics in
  Computing}} \bibinfo{volume}{2}, \bibinfo{number}{1} (\bibinfo{year}{2014}),
  \bibinfo{pages}{81--93}.
\newblock


\bibitem[Van~Bulck et~al\mbox{.}(2020)]%
        {vanbulck2020lvi}
\bibfield{author}{\bibinfo{person}{Jo Van~Bulck}, \bibinfo{person}{Daniel
  Moghimi}, \bibinfo{person}{Michael Schwarz}, \bibinfo{person}{Moritz Lipp},
  \bibinfo{person}{Marina Minkin}, \bibinfo{person}{Daniel Genkin},
  \bibinfo{person}{Yarom Yuval}, \bibinfo{person}{Berk Sunar},
  \bibinfo{person}{Daniel Gruss}, {and} \bibinfo{person}{Frank Piessens}.}
  \bibinfo{year}{2020}\natexlab{}.
\newblock \showarticletitle{{LVI: Hijacking Transient Execution through
  Microarchitectural Load Value Injection}}. In \bibinfo{booktitle}{\emph{41th
  IEEE Symposium on Security and Privacy (S\&P'20)}}.
\newblock


\bibitem[Xiao et~al\mbox{.}(2016)]%
        {xiao2016hardware}
\bibfield{author}{\bibinfo{person}{K Xiao}, \bibinfo{person}{D Forte},
  \bibinfo{person}{Y Jin}, \bibinfo{person}{R Karri}, \bibinfo{person}{S
  Bhunia}, {and} \bibinfo{person}{M Tehranipoor}.}
  \bibinfo{year}{2016}\natexlab{}.
\newblock \showarticletitle{Hardware Trojans: Lessons Learned after One Decade
  of Research}.
\newblock \bibinfo{journal}{\emph{ACM Transactions on Design Automation of
  Electronic Systems (TODAES)}} \bibinfo{volume}{22}, \bibinfo{number}{1}
  (\bibinfo{year}{2016}), \bibinfo{pages}{6}.
\newblock


\bibitem[Yang et~al\mbox{.}(2016)]%
        {yanga2}
\bibfield{author}{\bibinfo{person}{Kaiyuan Yang}, \bibinfo{person}{Matthew
  Hicks}, \bibinfo{person}{Qing Dong}, \bibinfo{person}{Todd Austin}, {and}
  \bibinfo{person}{Dennis Sylvester}.} \bibinfo{year}{2016}\natexlab{}.
\newblock \showarticletitle{{A2: Analog malicious hardware}}. In
  \bibinfo{booktitle}{\emph{{IEEE Symposium on Security and Privacy (S\&P)}}}.
  IEEE.
\newblock


\end{thebibliography}

\end{document}